# Nonmonotonic Scaling of the Anomalous Hall Effect in a Bicollinear Antiferromagnet

Ruifeng Wang[1,4], Chi Fang[1,4*], Ilya Kostanovski[1], Ke Xiao[1], Felix Küster[1], Jenny Davern[1], Naoto Nagaosa[2,3], Stuart S. P. Parkin[1*]

[1] Max Planck Institute of Microstructure Physics, Halle (Saale) 06120, Germany.

[2] RIKEN Center for Emergent Matter Science (CEMS), Wako, Saitama 351-0198, Japan.

[3] Fundamental Quantum Science Program (FQSP), TRIP Headquarters, RIKEN, Wako 351- 0198, Japan

[4] These authors contributed equally: Ruifeng Wang, Chi Fang

*Email: chi.fang@mpi-halle.mpg.de, stuart.parkin@mpi-halle.mpg.de

**Abstract**

An anomalous Hall effect (AHE) in antiferromagnetic (AF) systems with no net magnetization is of considerable interest for both fundamental physics and spintronic applications. Of particular interest is the two-dimensional van der Waals antiferromagnet FeTe that has an unusual fully magnetically compensated bicollinear AF structure and exhibits pronounced Kondo interaction leading to strong band renormalization. Here, we investigate the AHE in epitaxial FeTe thin films grown by molecular beam epitaxy. A large anomalous Hall conductivity is exhibited below the Néel temperature ($T_N$~60 K) and, strikingly, becomes nonlinear at high fields within a narrow temperature window around 49 K, deviating from conventional AHE scaling behavior versus its longitudinal conductivity. Linear fits reveal a pronounced negative peak in the intercept, accompanied by a field-induced canted magnetic moment. The AHE responses are related to the Berry curvature derived from FeTe's topological band structure, highlighting the intricate interplay between topology, magnetism, and electronic transport.

# Introduction

The anomalous Hall effect (AHE) has long been recognized as a fundamental transport phenomenon that reveals the intimate connection between magnetism and electronic properties in solids. In general, the AHE can originate from the extrinsic mechanism due to scattering from impurities and defects, such as the skew scattering and side jump, or from the intrinsic mechanism derived from the Berry curvature of the electronic bands (*1*). In ferromagnetic (FM) systems, both mechanisms usually coexist and, as a result, the intrinsic contribution to the AHE is often obscured and requires further distinction from extrinsic ones (*2*, *3*). This has encouraged extensive investigations into the magnitude scaling laws between anomalous Hall conductivity (AHC) $\sigma_{xy}$ and longitudinal conductivity $\sigma_{xx}$ in a variety of ferromagnetic materials. Typically, $\sigma_{xy} \propto \sigma_{xx}$ for skew scattering and $\sigma_{xy} \propto constant$ for side jump or the intrinsic mechanism. Together, these yield a monotonic conductivity dependence below the magnetic ordering temperature and are well explained by quantum transport theory (*4*). Meanwhile, AHE has also been reported in antiferromagnetic (AF) systems (*5–11*). The vanishing net magnetization suppresses extrinsic scattering channels, while broken symmetries may generate large Berry curvature, making it possible to realize an AHE that is dominantly intrinsic in origin. As a result, the AHE in AF systems fails to be reliably inferred from the magnetization. Instead, the developing scaling analysis could provide a framework for understanding the coexistence of AHE mechanisms in AF materials (*12–14*).

From another perspective, deviations from the conventional scaling relation probably indicates the involvement of additional mechanisms. An obvious breakdown of scaling laws is reported in a ferromagnetic USbTe with strong electron correlation (*15*). A Kondo lattice, where the array of local moments interacts coherently with the conduction electrons via an antiferromagnetic interaction (*16*), can generate flat bands and thereby effectively modify the Berry curvature as the temperature varies. The extrinsic mechanism, in addition, is also in competition, making the analysis of scaling relation more complicated (*15*). Recently, Kondo lattice behavior is found in AF systems, e.g., $CeTe_3$ (*17*), $CeIn_3$ (*18*), UOTe (*19*), $Mn_{3+x}Sn_{1-x}$ (*20*), and FeTe (*21*). But the AHE and its scaling relations in these systems are not fully revealed yet. Among them, tetragonal FeTe represents a particularly intriguing case (*21*, *22*) with a high Kondo temperature (*21*). It is a metallic vdW material that becomes monoclinic below its Néel temperature, $T_N$, and hosts a fully compensated bicollinear AF (BCAF) order (*23*, *24*). At the same time, FeTe possesses a topologically non-trivial band structure featuring nodal lines and nodal points near the Fermi energy (*25*, *26*), which could give rise to large Berry curvature. Nevertheless, the intrinsic AHE is always missing or ambiguous in the BCAF order, as the imperfections in the layers (*27*, *28*) or interstitial Fe atoms (*29*) may generate a net magnetization, leading to a ferromagnet-like AHE in low magnetic fields (<2 T). The relationship between the BCAF order and the topological band structure governing the AHE intrinsically in FeTe has not yet been established.

Here, we report the observation of an AHE, driven by the intrinsic Berry curvature, and its nonmonotonic scaling relation in epitaxial films of BCAF FeTe in the presence of a magnetic field. High-quality single-crystal FeTe thin films were prepared by molecular beam epitaxy on (001) SrTiO$_3$ single crystalline substrates. These films exhibit a Néel temperature ($T_N$) of ~60 K similar to that reported in bulk single crystals. They exhibit a BCAF order below $T_N$ (Fig. 1, A and B) which arises from a competition between nearest, 2$^{nd}$ and 3$^{rd}$ neighbor magnetic exchange interactions among Fe atoms that form a face-centered square lattice within a single plane (*30*, *31*). These planes are separated from one another by planes of Te atoms which themselves are separated from each other by the vdW gap. Above $T_N$, tetragonal FeTe exhibits a Hall conductivity (HC) that is proportional to the magnetic field with a positive slope. However, in a specific temperature range just below $T_N$, the slope of the HC changes sign, accompanied by the emergence of a nonlinear AHC at high fields. These results highlight a remarkable instance of significant AHE governed by intrinsic mechanisms in a collinear AF.

**Results**

**Preparation of single-crystal bicollinear FeTe films**

High-quality FeTe films were prepared and precisely controlled using a state-of-the-art VEECO molecular beam epitaxy system (see Materials and Methods). Under optimized growth parameters, FeTe films were epitaxially grown on SrTiO$_3$ (001) (STO) substrates in the Volmer-Weber (VW) mode, as evidenced by sharp reflection high-energy electron diffraction (RHEED) patterns and clear surface atomic force microscopic images (Fig. S1). The surface topography of FeTe films (Fig. S1, B-E) reveals uniformly oriented tetragonal islands coalesced into a continuous thin film that fully covers the STO substrate. Further analysis highlights well-defined terrace edges with a pronounced fourfold symmetry, indicative of the high single-crystal quality of the FeTe film in the tetragonal structure. More importantly, the BCAF order exists in a narrow growth window (*32–35*). Rutherford backscattering spectrometry (RBS) revealed the stoichiometry of our films to be $Fe_{1.08}Te$ (Fig. S2), consistent with results on bulk $Fe_{1+x}Te$ (*32*, *33*), which only forms and maintains BCAF ordering within the composition range $0.05 < x < 0.12$. Fig. 1C shows an atomically resolved scanning tunneling microscopy (STM) topography of the FeTe film. In contrast to cleaved bulk FeTe (*34*, *35*), our film exhibits a defect-free surface without excess Fe atoms. Using a spin-polarized tip, the STM image reveals a striped magnetic contrast with a spacing of $2a$ (Fig. 1D), consistent with the BCAF ordering observed in FeTe with low excess Fe content (*34*, *35*). The films' exceptional crystallinity was further corroborated by x-ray diffraction (XRD) analysis (Fig. 1E) and scanning transmission microscopy (STEM) imaging (Fig. 1F). The STEM cross-sectional image resolved atomic-scale structure at the FeTe/STO interface, while XRD patterns revealed a series of (00$n$) Bragg peaks, satisfying the condition $2d \sin\theta = n\lambda$, where $\lambda = 1.5406\,\text{Å}$ is the x-ray wavelength. From these measurements, the lattice parameters of the FeTe film were determined to be $a = b = 3.84\,\text{Å}$, $c = 6.26\,\text{Å}$, consistent with expectations for this material. Notably, the FeTe/STO interfacial structure, which exhibits a distinct $SrO$-$Ti_xO_2$-$Ti_xO_2$-Te-Fe-Te stacking sequence, as illustrated in the atomic model superimposed on Fig. 1F. The observation of the double titanium oxide termination at the STO interface is noteworthy, as this feature has been extensively characterized in the analogous FeSe/STO interface (*36*, *37*), but there have been no reports to date in the FeTe/STO system.

**Anomalous Hall conductivity in FeTe**

Electrical measurements disclose the role played by the BCAF order in FeTe. Longitudinal resistivity ($\rho_{xx}$) measurements using a four-probe van der Pauw method (see Materials and Methods and Fig. S4, A-C) reveal a pronounced maximum at 60 K (Fig. 2A), which is a clear signature of the phase transition from paramagnetic (PM) to AF order at $T_N$. The magnetic order transition is accompanied by a structural phase transition from tetragonal to monoclinic which stabilizes the BCAF order (*23*, *24*). The peak in the $\rho_{xx}$-$T$ curve is likely driven by enhanced spin fluctuations, as observed for other AF orderings (*38*, *39*). Above $T_N$, $\rho_{xx}$ exhibits a logarithmic temperature dependence, $\rho(T) = a + b \cdot \ln(T)$ (fitting details are provided in Fig. S5), consistent with the behavior previously reported in FeTe single crystals with Kondo interactions (*21*). This logarithmic dependence reflects the interplay between localized spin fluctuations and itinerant carriers in the PM regime. Below $T_N$, $\rho_{xx}$ decreases with decreasing temperature, exhibiting metallic behavior indicative of suppressed scattering due to the onset of the AF order and Kondo-lattice-like coherence (*21*). Intriguingly, a low-temperature upturn appears below 8 K, persisting to the lowest measurement temperature of 2 K. To elucidate this behavior, the $\rho_{xx}(T)$ data below 30 K were fitted using a model incorporating multiple scattering mechanisms: $\rho(T) = a + b \cdot T^2 + c \cdot \ln(T) + d \cdot T^5$. Here, the $T^2$ term describes Fermi liquid behavior, while the $T^5$ term accounts for phonon-induced scattering, although its contribution is negligible compared to the other contributions. The $\ln(T)$ term, indicative of Kondo-like behavior, suggests interactions between conduction electrons and excess Fe atoms, highlighting the complex low-temperature electronic landscape of FeTe.

To explore the AHE in FeTe, transverse resistivity ($\rho_{xy}$) measurements in an out-of-plane magnetic field (*B*) up to 13.5 T were performed from 12 to 300 K (Fig. 2B). At low fields (< 2 T), $\rho_{xy}(B)$ follows a linear dependence on *B*, resembling an OHE-like behavior in both the AF and PM states, consistent with the literature (*27, 29*). However, at high magnetic fields, an unexpected Hall response, which is non-proportional to *B*, emerges, becoming particularly prominent in the intermediate temperature range (40 to 55 K). This Hall response is quantified by subtracting a linear background by fitting the high-field region (*B* > 12.5 T) of the $\rho_{xy}$ - *B* curves (Fig. 2C). The intercept of these fitted curves defines the magnitude of the residual Hall resistivity ($\rho_{xy}^r$), which also serves as a measure of the curve's nonlinearity. In general, $\rho_{xy}^r$ contains both any intrinsic as well as any extrinsic AHE, the latter of which is usually proportional to the net magnetization (*M*). However, the similar nonlinear field dependence of the magnetization is absent, as discussed later, ruling out a dominant extrinsic contribution to $\rho_{xy}^r$. As shown in Fig. 2D and its inset, both $\rho_{xy}^r(T)$ and the corresponding Hall conductivity, $\sigma_{xy}^r(T) \approx \rho_{xy}^r(T) / \rho_{xx}^2(T)$, exhibit a striking negative peak within a narrow temperature window around 49 K, noted as $T_p$, while remaining nearly zero at temperatures away from it. $\sigma_{xy}^r$ increases by almost 10 times from 12 K to 49 K and then rapidly shrinks to nearly zero beyond $T_N$. To further elucidate the origin of this peak, we inspect the scaling relation of $\rho_{xy}^r$ vs. $\rho_{xx}$ and absolute value $|\sigma_{xy}^r|$ vs. $\sigma_{xx} = 1/\rho_{xx}$, respectively. As shown in Fig. 2E, $\rho_{xy}^r$ is negligible and shows no clear correlation with $\rho_{xx}$ above $T_N$. Below $T_N$, $\rho_{xy}^r$ maintains an obvious nonmonotonic scaling relation with $\rho_{xx}$, reaching a pronounced minimum near $\rho_{xx}$ = 400 μΩ·cm, which corresponds to $T_p$. We use a logarithmic scale to present the scaling of the $|\sigma_{xy}^r|$ vs. $\sigma_{xx}$ for $T \leq T_N$ in Fig.

2F. The correlation gradually turned from positive to negative over $T_p$. Although the positive section might relate to the scattering process, the negative section lacks of an intuitive origin. Such deviation from the conventional scaling law behavior suggests that the underlying AHE mechanism is not dominated by ordinary mechanisms alone.

Several possible mechanisms could be accountable. We first rule out the non-AHE ones. A nonlinear field dependence of the OHE has often been reported in high mobility ($\mu_i$, $i = e$ for electrons or $h$ for holes) systems (*40, 41*). Such behavior arises from the coexistence of electrons and holes, where the effective Hall conductivity is tuned by their relative concentrations and mobilities within a multi-band framework. To consider this, we extracted the slopes at low field, $R_H^{Low}$, and high field, $R_H^{High}$, as shown in Figs. 3A and B. Pronounced nonlinearities in the field dependence are expected to emerge only when the cyclotron motion of the carriers becomes sufficiently strong. Conversely, when $\omega_c\tau \ll 1$, the OHE coefficient is constant and gives only a linear field dependence of $\rho_{xy}$, where $\omega_c$ is the cyclotron frequency and $\tau$ is the relaxation time (*42, 43*). We find that the carrier mobility $\mu$ in our films, estimated from $R_H^{High}$ (Fig. S4D), is very low (comparable to values found in previous work (*28*)). The magnitude of $\mu$ satisfies the low field criterion, $\omega_c\tau = \mu B \sim 10^{-3} \ll 1$. Therefore, a nonlinear OHE response is not expected in our films. Additional insight is provided by the longitudinal thermoelectric response, i.e., the Seebeck effect which indirectly reflects the carrier types and mobilities. We performed thermoelectric voltage ($V_{th}$) measurements using the setup shown in Fig. 3C, and compared the results with the nominal Hall coefficient. The drift of carriers under a temperature gradient generates a voltage $V_{th}$, whose sign and magnitude reflect the dominant carrier type and its relative density. As displayed in Fig. 3D, the measured $V_{th}$ exhibits two distinct sign reversals, signaling carrier-type transitions. The first reversal, from positive to negative around 45 K, marks a switch from electron to hole dominance. This behavior is consistent with the sign change in the $R_H^{High}$ (Fig. 3B). Taken together, both the low mobility and the thermoelectric data indicate that the nonlinearities in $\rho_{xy}^r$ of FeTe cannot be explained by multiband OHE and point towards the AHE mechanism underlying the observed nonlinear Hall response.

**Tiny magnetic moment in FeTe under an external field**

The magnitude of AHE can be tuned by the magnetization, which affects the spin polarization, especially the out-of-plane component in our case. Next, we examine the role played by the magnetic-field-induced magnetization. We measured the field dependence of $M$ in our films using a magnetometer (see Materials and Methods). The $M(B)$ curves, shown in Fig. 4A for out-of-plane field ($B//c$) and in Fig. S6 for in-plane field (B//$a$), are dominated by the diamagnetic signal from the STO substrate (Fig. S7). The extracted nonlinear contribution (shown in Fig. 4B) reveals a very small $M$ with no obvious temperature dependence. The field dependence in Fig. 4B is clearly different from that of $\rho_{xy}$ (Fig. 2C). The saturation magnetization ($M_S$) extracted from the high-field regions remains very small across the entire temperature range (Fig. 4C middle panel). The contribution from the STO substrate (Fig. 4C bottom panel) was measured on a bare substrate and subtracted from the total signal. The isolated contribution from the FeTe film (Fig. 4C top panel) shows only a very small magnetic moment without any significant temperature dependence unlike that of $\rho_{xy}^r(T)$ in Fig 2D. Specifically, $\rho_{xy}^r$ in our case emerges near 49 K, rather than at lower temperatures, and shows no correlation with the measured $M$. Nevertheless, the $M(B)$ loops do indicate the possibility of small canted magnetic moments with a very small tilt angle about 0.26°, which breaks time reversal symmetry and could generate an intrinsic AHE. We further carried out polar reflective magnetic circular dichroism (RMCD) measurements to probe the field dependence of the $M_{FeTe}$ along the out-of-plane direction (Fig. 4D). The RMCD signal arises from within the limited skin depth and therefore is more sensitive to the FeTe moment. The signal varies only linearly with field with a positive slope at each temperature. A clear peak in this slope is observed near $T_N$ that is characteristic of an AF order. While a simple extrinsic AHE from spin canting along hard-axis would give rise to a linear field-dependent contribution to $\rho_{xy}$, the nonlinear field-dependent contribution $\rho_{xy}^r$ can arise intrinsically from the evolution of the electronic structure in magnetic field. For the AF FeTe system, in the absence of an external field, the Berry curvature vanishes due to the time-reversal symmetry above $T_N$ and the combined symmetry of time-reversal and spatial inversion below $T_N$, which enforces a two-fold Kramers' degeneracy across all bands. However, applying a magnetic field induces a net $M$ due to spin canting and breaks such a symmetry, lifting band degeneracies and opening a gap at certain symmetry-protected band crossings, depending on the details of the band structure and symmetry protection. This induces a Berry curvature redistribution in momentum space and generates a nonzero Berry curvature integral below the Fermi surface.

**Mechanism analysis**

The $\rho_{xy}^r$ is likely of an intrinsic origin with an unusual temperature dependence and scaling relation. The AHE appears as the BCAF builds up and the negative maximum of $\rho_{xy}^r$ takes place at $T_p$ just below $T_N$. The PM to BCAF phase change should be responsible. However, a spin fluctuation mechanism, which could result in nonmonotonic $T$ dependence, cannot explain the scaling of $\rho_{xy}^r$ in our case as it would rather give rise to a maximum in the AHC at $T_N$. The carrier type change is also not a good candidate as its crossover temperature is around 45 K, a little smaller than the $T_p$. Alternatively, we come to the Kondo physics aspect. We show the temperature dependence of the first derivative of $\rho_{xx}$ with respect to $T$, i.e., $d\rho_{xx}/dT$, in Fig. 5A, which is thought to be sensitive to phase changes. Simultaneously, as the BCAF order emerges as temperature decreases, the spin fluctuation is fading and coherent Kondo scattering is forming to create a Kondo lattice, which leads to the decrease of the $\rho_{xx}$ (*16*). The local hybridization between localized Fe $3d_{xy}$ and itinerant Te $5p_z$ orbitals is thought to reconstruct the band structure below $T_N$ (*21*). A peak at 48.3 K indicates the dramatic evolution of the band structure, which is, surprisingly, almost the same as $T_p$. We show that the $\rho_{xy}^r$ exhibits a strong positive correlation with $d\rho_{xx}/dT$ (Fig. 5B). Temperature alone, via the broadening of the Fermi–Dirac distribution, is unlikely to produce such a large effect of AHE. Instead, it more likely drives a reconstruction of the electronic bands by renormalizing the strength of exchange interactions. Specifically, the coherent Kondo interaction drives band renormalization which enhances Berry curvature near the $T_p$, leading to a strongly temperature-dependent net Berry-curvature contribution to the AHE.

**Discussion**

Compared to other collinear AF systems, the observed magnitude of the AHE is high in FeTe (*5*). However, the magnitude is at least one order of magnitude smaller than those due to an intrinsic mechanism in non-collinear AF materials (*5*) or in some FMs (*1*). This could arise from several factors. While FeTe also has domains with different directions of the BCAF ordering, all domains have, in a nonzero magnetic field, a small net magnetic moment induced along the field direction from spin canting. Thus, contributions from different domains do not cancel each other and the AHC remains robust against variations in the AF domain distribution. However, since the present measurements are performed with the current confined to the (001) plane, we cannot rule out the possibility that FeTe exhibits different AHE responses with current along other directions, especially for current applied perpendicular to the vdW gap. Similar to the case of non-collinear AF ones (*7, 8*), a single magnetic domain in FeTe may host a spontaneous nonzero AHE when the current is applied along the [001] direction. Meanwhile, spin disorder at magnetic and structural domain walls (*44*) can alter the total AHC. The relationship between the AHE and the magnetic domain structures in FeTe remains to be further investigated, as the domain configuration is, in practice, difficult to control. Also, although fluctuation effects cannot account for the AHE origin in FeTe, they could modify the AHC magnitude (*8*, *45*).

Overall, the combination of magnetization, Hall, and thermoelectric measurements provides a coherent picture: The Hall response in FeTe/STO films is primarily governed by the intrinsic topology of the band structure. The absence of significant magnetic moments, the temperature evolution of the AHC, and the Berry curvature distribution all point to a topological origin of the AHE, providing crucial insights into the interplay between magnetism and electronic band topology in correlated systems. Our findings establish FeTe as a compelling platform for probing intrinsic topological phenomena in antiferromagnets. By elucidating the interplay between magnetism, topology, and electronic transport, this work not only deepens our understanding of the Hall effect in collinear AF systems but also opens avenues for future exploration of Berry curvature-driven phenomena, with significant potential applications in spintronic technologies.

**Materials and Methods**

**Growth of FeTe/SrTiO$_3$ films**

FeTe films in this work were grown on SrTiO$_3$(STO) (100) substrates using a Veeco GEN10 MBE system under ultrahigh vacuum (UHV) conditions better than $2\times10^{-10}$ Torr to minimize impurities. Prior to growth, the STO substrates underwent chemical treatment and high-temperature annealing to achieve an atomically smooth, TiO$_2$-terminated surface. The cleaning process involved ultrasonic treatment in acetone and isopropanol to remove adsorbed impurities, followed by immersion in a 90°C hot water bath for 20 min to convert surface SrO into hydroxide complexes. These complexes were then dissolved in a 9 wt% HCl solution for 20 min. After thorough rinsing with deionized water, the substrates were annealed at 1000°C for 3 hours in a tube furnace under an oxygen atmosphere, resulting in well-defined TiO$_2$ step-terrace structures. Once transferred to the UHV chamber, they were further degassed at 600°C for 2 hours before film deposition.

FeTe films were synthesized by co-depositing Fe (99.99% purity) via a high-temperature Knudsen cell and Te (99.999% purity) via a valved cracker cell, with the substrate temperature maintained at 350°C. Beam flux was characterized using a beam flux monitor, ensuring a Te:Fe flux ratio of approximately 9.5:1 and a growth rate of 1 ML/min. Real-time RHEED monitoring (kSA 400) was employed to track surface evolution. After deposition, the films were annealed at the growth temperature for 20 min to remove excess Te and enhance crystallinity. Finally, a ~1 nm Al capping layer was deposited in situ after the sample cooled to room temperature. Upon air exposure, this layer naturally oxidized into a dense $AlO_x$ barrier, effectively preventing FeTe oxidation.

X-ray photoelectron spectroscopy (XPS) further indicated the binding energies of Fe and Te remained in their $Fe^0$ and $Te^0$ states with the protective $Al_2O_3$ capping layer (Fig. S3), ruling out surface oxidation or degradation into other impurity compounds (*46*). Unless otherwise specified, all data presented in this work were obtained from FeTe films with a uniform thickness of 25 monolayers (MLs).

**Structural and magnetism characterization**

The structure and quality of the FeTe/STO films were characterized using X-ray diffraction (XRD), atomic force microscopy (AFM), scanning transmission electron microscopy (STEM), and scanning tunneling microscopy (STM). XRD measurements were performed on a high-resolution diffractometer (Bruker D8) equipped with monochromatic Cu K$\alpha_1$ radiation ($\lambda$ = 1.5406 Å), operating at 40 kV and 40 mA. Surface topography was examined by AFM (Bruker Dimension Icon) under ambient conditions. Measurements were conducted in tapping mode with a scan rate of 1 Hz. The acquired images were processed with NanoScope Analysis software using standard flattening procedures. High-angle annular dark-field (HAADF) STEM images were acquired on a spherical-aberration-corrected transmission electron microscope. Atomically resolved surface structure and magnetic ordering were investigated using an UHV STM apparatus (Unisoku USM1600) at 5 K. The FeTe sample for STM measurements was grown on a 0.5 wt% Nb-doped STO substrate using the same recipe as for the transport samples in the Veeco chamber. To prevent surface degradation during the rapid ex situ transfer, a 30-nm-thick Te capping layer was deposited at room temperature and subsequently removed by annealing at 350°C for 30 min in the STM chamber. Topographic images were acquired with a Cr bulk tip, and magnetic contrast was obtained using a spin-polarized tip prepared by gently indenting the tip into the surface to attach a Fe cluster (*47*). The STM images were analyzed by fast Fourier transform (FFT) using WSxM software. The magnetic moment was determined using a QD MPMS3 as the mentioned magnetometer. RMCD was measured with QD OptiCool.

**Transport measurements**

Electronical transport measurements were conducted using a Physical Property Measurement System (QD DynaCool 1.9 K, 14 T) via the van der Pauw (vdP) method unless otherwise specified. Four electrodes were manually bonded at the corners of a square sample, as illustrated in Fig. S4A, using a wire bonder (West bond). For longitudinal resistivity ($\rho_{xx}$) measurements, a direct current ($I_{12} = 20$ µA) was applied from electrode 1 to electrode 2 using a Keithley 6221 current source, and the voltage ($U_{43}$, defined as $U_4 - U_3$) was measured between electrodes 3 and 4 using a Keithley 2182A voltmeter. The $\rho_{xx}$ was then calculated by averaging the resistances in two orthogonal directions, according to the formula: $\rho_{xx} = \frac{\pi}{\ln 2} \cdot \frac{U_{43}/I_{12} + U_{23}/I_{14}}{2} \cdot t$, where $\frac{\pi}{\ln 2}$ is the vdP constant and $t$ is the film thickness. The corresponding longitudinal conductivity was determined by $\sigma_{xx} = 1/\rho_{xx}$.

For Hall effect measurements, a direct current ($I_{13} = 100$ µA) was applied along one diagonal direction of the sample while the voltage ($U_{42}$) was measured along the orthogonal diagonal. The transverse resistivity ($\rho_{xy}$) was calculated using: $\rho_{xy} = U_{42}/I_{13} \cdot t$. To correct for any asymmetry in electrode placement, the $\rho_{xy}$ at each magnetic field strength was antisymmetrized using: $\rho_{xy}(B) = \frac{\rho_{xy}(B) - \rho_{xy}(-B)}{2}$. Subsequently, the Hall conductivity was determined by $\sigma_{xy}(B,T) = \frac{\rho_{xy}(B,T)}{\rho_{xx}(0,T)^2}$, using the zero-field value of $\rho_{xx}$, which was found to be negligibly dependent on magnetic field compared to $\rho_{xy}$ (Fig. S4, B and C). All the temperature-dependent measurements are conducted as warming up.

# References

1. N. Nagaosa, J. Sinova, S. Onoda, A. H. MacDonald, N. P. Ong, Anomalous Hall effect. *Rev. Mod. Phys.* **82**, 1539–1592 (2010).

2. S. Onoda, Intrinsic versus extrinsic anomalous Hall effect in ferromagnets. *Phys. Rev. Lett.* **97** (2006).

3. Y. Tian, L. Ye, X. Jin, Proper scaling of the anomalous Hall effect . *Phys. Rev. Lett.* **103**, 087206 (2009).

4. S. Onoda, Quantum transport theory of anomalous electric, thermoelectric, and thermal Hall effects in ferromagnets. *Phys. Rev. B* **77** (2008).

5. L. Šmejkal, A. H. MacDonald, J. Sinova, S. Nakatsuji, T. Jungwirth, Anomalous Hall antiferromagnets. *Nat. Rev. Mater.* **7**, 482–496 (2022).

6. R. Shindou, N. Nagaosa, Orbital ferromagnetism and anomalous Hall effect in antiferromagnets on the distorted fcc lattice. *Phys. Rev. Lett.* **87**, 116801 (2001).

7. V. Baltz, A. Manchon, M. Tsoi, T. Moriyama, T. Ono, Y. Tserkovnyak, Antiferromagnetic spintronics. *Rev. Mod. Phys.* **90**, 015005 (2018).

8. A. K. Nayak, J. E. Fischer, Y. Sun, B. Yan, J. Karel, A. C. Komarek, C. Shekhar, N. Kumar, W. Schnelle, J. Kübler, C. Felser, S. S. P. Parkin, Large anomalous Hall effect driven by a nonvanishing Berry curvature in the noncolinear antiferromagnet $Mn_3Ge$. *Sci. Adv.* **2**, e1501870 (2016).

9. S. Nakatsuji, N. Kiyohara, T. Higo, Large anomalous Hall effect in a non-collinear antiferromagnet at room temperature. *Nature* **527**, 212–215 (2015).

10. X. Li, J. Koo, Z. Zhu, K. Behnia, B. Yan, Field-linear anomalous Hall effect and Berry curvature induced by spin chirality in the kagome antiferromagnet $Mn_3Sn$. *Nat. Commun.* **14**, 1642 (2023).

11. D. Zhu, J. Lu, Y. Jiang, Z. Zheng, D. Wang, C. Zhou, J. Zhou, S. Chen, Y. Gu, L. Liu, P. Yang, K. Shi, S. Peng, G. Xing, W. Zhao, J. Chen, Observation of anomalous Hall effect in collinear antiferromagnet IrMn. *Nano Lett.* **25**, 4307–4313 (2025).

12. S. Xu, B. Dai, Y. Jiang, D. Xiong, H. Cheng, L. Tai, M. Tang, Y. Sun, Y. He, B. Yang, Y. Peng, K. L. Wang, W. Zhao, Universal scaling law for chiral antiferromagnetism. *Nat. Commun.* **15**, 3717 (2024).

13. H. Jeon, H. Seo, J. Seo, Y. H. Kim, E. S. Choi, Y. Jo, H. N. Lee, J. M. Ok, J. S. Kim, Large anomalous Hall conductivity induced by spin chirality fluctuation in an ultraclean frustrated antiferromagnet $PdCrO_2$. *Commun. Phys.* **7**, 162 (2024).

14. L. Song, F. Zhou, H. Li, B. Ding, X. Li, X. Xi, Y. Yao, Y.-C. Lau, W. Wang, Large anomalous Hall effect at room temperature in a Fermi-level-tuned Kagome antiferromagnet. *Adv. Funct. Mater.* **34**, 2316588 (2024).

15. H. Siddiquee, C. Broyles, E. Kotta, S. Liu, S. Peng, T. Kong, B. Kang, Q. Zhu, Y. Lee, L. Ke, H. Weng, J. D. Denlinger, L. A. Wray, S. Ran, Breakdown of the scaling relation of anomalous Hall effect in Kondo lattice ferromagnet USbTe. *Nat. Commun.* **14**, 527 (2023).

16. P. Coleman, *Introduction to Many-Body Physics* (Cambridge University Press, 2015).

17. H. Zeng, Y. Zhang, B. Ji, J. Cai, S. Zou, Z. Wang, C. Dong, K. Luo, Y. Yuan, K. Wang, J. Zhang, C. Xi, J. Wang, L. Li, Y. Dai, J. Li, Y. Luo, Kondo-coupled van der Waals antiferromagnet with high-mobility quasiparticles. *Newton* **2**, 100320 (2026).

18. W. Simeth, Z. Wang, E. A. Ghioldi, D. M. Fobes, A. Podlesnyak, N. H. Sung, E. D. Bauer, J. Lass, S. Flury, J. Vonka, D. G. Mazzone, C. Niedermayer, Y. Nomura, R. Arita, C. D. Batista, F. Ronning, M. Janoschek, A microscopic Kondo lattice model for the heavy fermion antiferromagnet $CeIn_3$. *Nat. Commun.* **14**, 8239 (2023).

19. C. Broyles, S. Mardanya, M. Liu, J. Ahn, T. Dinh, G. Alqasseri, J. Garner, Z. Rehfuss, K. Guo, J. Zhu, D. Martinez, D. Li, Y. Hao, H. Cao, M. Boswell, W. Xie, J. G. Philbrick, T. Kong, L. Yang, A. Vishwanath, P. Kim, S.-Y. Xu, J. E. Hoffman, J. D. Denlinger, S. Chowdhury, S. Ran, UOTe: Kondo-interacting topological antiferromagnet in a van der waals lattice. *Adv. Mater.* **37**, 2414966 (2025).

20. D. Khadka, T. R. Thapaliya, S. Hurtado Parra, X. Han, J. Wen, R. F. Need, P. Khanal, W. Wang, J. Zang, J. M. Kikkawa, L. Wu, S. X. Huang, Kondo physics in antiferromagnetic Weyl semimetal $Mn_{3+x}Sn_{1-x}$ films. *Sci. Adv.* **6**, eabc1977 (2020).

21. Y. Kim, M.-S. Kim, D. Kim, M. Kim, M. Kim, C.-M. Cheng, J. Choi, S. Jung, D. Lu, J. H. Kim, S. Cho, D. Song, D. Oh, L. Yu, Y. J. Choi, H.-D. Kim, J. H. Han, Y. Jo, J. H. Shim, J. Seo, S. Huh, C. Kim, Kondo interaction in FeTe and its potential role in the magnetic order. *Nat. Commun.* **14**, 4145 (2023).

22. L. Kang, C. Ye, X. Zhao, X. Zhou, J. Hu, Q. Li, D. Liu, C. M. Das, J. Yang, D. Hu, J. Chen, X. Cao, Y. Zhang, M. Xu, J. Di, D. Tian, P. Song, G. Kutty, Q. Zeng, Q. Fu, Y. Deng, J. Zhou, A. Ariando, F. Miao, G. Hong, Y. Huang, S. J. Pennycook, K.-T. Yong, W. Ji, X. Renshaw Wang, Z. Liu, Phase-controllable growth of ultrathin 2D magnetic FeTe crystals. *Nat. Commun.* **11**, 3729 (2020).

23. E. E. Rodriguez, C. Stock, P. Zajdel, K. L. Krycka, C. F. Majkrzak, P. Zavalij, M. A. Green, Magnetic-crystallographic phase diagram of the superconducting parent compound $Fe_{1+x}Te$. *Phys. Rev. B* **84**, 064403 (2011).

24. W. Bao, Tunable ($\delta\pi$, $\delta\pi$)-type antiferromagnetic order in $\alpha$-Fe(Te,Se) superconductors. *Phys. Rev. Lett.* **102** (2009).

25. M. G. Vergniory, L. Elcoro, C. Felser, N. Regnault, B. A. Bernevig, Z. Wang, A complete catalogue of high-quality topological materials. *Nature* **566**, 480–485 (2019).

26. Topological Materials Database, https://topologicalquantumchemistry.org.

27. I. Tsukada, M. Hanawa, S. Komiya, A. Ichinose, T. Akiike, Y. Imai, A. Maeda, Mobility analysis of FeTe thin films . *J. Phys. Soc. Jpn.* **80**, 023712 (2011).

28. I. Tsukada, M. Hanawa, S. Komiya, A. Ichinose, T. Akiike, Y. Imai, A. Maeda, Hall effect of FeTe and Fe(Se$_{1-x}$Te$_x$) thin films. *Physica C: Superconductivity and its Applications* **471**, 625–629 (2011).

29. M. Meng, S. Liu, D. Song, X. Zhang, H. Du, H. Huang, H. Liu, Z. Sun, C. Mei, H. Yang, H. Tian, Y. Lu, Y. Zhang, J. Li, Y. Zhao, Magnetotransport property of oxygen-annealed Fe$_{1+y}$Te thin films. *J. Phys.: Condens. Matter* **35**, 305701 (2023).

30. F. Ma, W. Ji, J. Hu, Z.-Y. Lu, T. Xiang, First-principles calculations of the electronic structure of tetragonal α-FeTe and α-FeSe Crystals: Evidence for a bicollinear antiferromagnetic order. *Phys. Rev. Lett.* **102**, 177003 (2009).

31. M. J. Han, S. Y. Savrasov, Doping driven ($\pi$, 0) nesting and magnetic properties of Fe$_{1+x}$Te superconductors. *Phys. Rev. Lett.* **103**, 067001 (2009).

32. S. Rößler, D. Cherian, W. Lorenz, M. Doerr, C. Koz, C. Curfs, Yu. Prots, U. K. Rößler, U. Schwarz, S. Elizabeth, S. Wirth, First-order structural transition in the magnetically ordered phase of Fe$_{1.13}$Te. *Phys. Rev. B* **84**, 174506 (2011).

33. C. Koz, S. Rößler, A. A. Tsirlin, S. Wirth, U. Schwarz, Low-temperature phase diagram of Fe$_{1+y}$Te studied using x-ray diffraction. *Phys. Rev. B* **88**, 094509 (2013).

34. M. Enayat, Z. Sun, U. R. Singh, R. Aluru, S. Schmaus, A. Yaresko, Y. Liu, C. Lin, V. Tsurkan, A. Loidl, J. Deisenhofer, P. Wahl, Real-space imaging of the atomic-scale magnetic structure of Fe$_{1+y}$Te. *Science* **345**, 653–656 (2014).

35. C. Trainer, C. M. Yim, C. Heil, F. Giustino, D. Croitori, V. Tsurkan, A. Loidl, E. E. Rodriguez, C. Stock, P. Wahl, Manipulating surface magnetic order in iron telluride. *Sci. Adv.* **5**, eaav3478 (2019).

36. F. Li, Q. Zhang, C. Tang, C. Liu, J. Shi, C. Nie, G. Zhou, Z. Li, W. Zhang, C.-L. Song, K. He, S. Ji, S. Zhang, L. Gu, L. Wang, X.-C. Ma, Q.-K. Xue, Atomically resolved FeSe/SrTiO$_3$(001) interface structure by scanning transmission electron microscopy. *2D Mater.* **3**, 024002 (2016).

37. H. Sims, D. N. Leonard, A. Y. Birenbaum, Z. Ge, T. Berlijn, L. Li, V. R. Cooper, M. F. Chisholm, S. T. Pantelides, Intrinsic interfacial van der Waals monolayers and their effect on the high-temperature superconductor FeSe/SrTiO$_3$. *Phys. Rev. B* **100**, 144103 (2019).

38. P. E. Brommer, N. H. Duc, "Magnetoelasticity in Nanoscale Heterogeneous Materials" in *Encyclopedia of Materials: Science and Technology*, K. H. J. Buschow, R. W. Cahn, M. C. Flemings, B. Ilschner, E. J. Kramer, S. Mahajan, P. Veyssière, Eds. (Elsevier, Oxford, 2004).

39. C. Fang, C. Wan, X. Zhang, S. Okamoto, T. Ma, J. Qin, X. Wang, C. Guo, J. Dong, G. Yu, Z. Wen, N. Tang, S. S. P. Parkin, N. Nagaosa, Y. Lu, X. Han, Observation of the fluctuation spin Hall effect in a low-resistivity antiferromagnet. *Nano Lett.* **23**, 11485–11492 (2023).

40. K. Wang, C. Petrovic, Multiband effects and possible Dirac states in LaAgSb$_2$. *Phys. Rev. B* **86**, 155213 (2012).

41. J. E. Dill, C. F. C. Chang, D. Jena, H. G. Xing, Two-carrier model-fitting of Hall effect in semiconductors with dual-band occupation: A case study in GaN two-dimensional hole gas. *J. Appl. Phys.* **137**, 025702 (2025).

42. C. M. Hurd, *The Hall Effect in Metals and Alloys*, Ch. 2 (Springer US, Boston, MA, 1972).

43. J. Sólyom, *Fundamentals of the Physics of Solids*, Ch. 24 (Springer, 2011).

44. O. J. Amin, A. Dal Din, E. Golias, Y. Niu, A. Zakharov, S. C. Fromage, C. J. B. Fields, S. L. Heywood, R. B. Cousins, F. Maccherozzi, J. Krempaský, J. H. Dil, D. Kriegner, B. Kiraly, R. P. Campion, A. W. Rushforth, K. W. Edmonds, S. S. Dhesi, L. Šmejkal, T. Jungwirth, P. Wadley, Nanoscale imaging and control of altermagnetism in MnTe. *Nature* **636**, 348–353 (2024).

45. Z. Fang, N. Nagaosa, K. S. Takahashi, A. Asamitsu, R. Mathieu, T. Ogasawara, H. Yamada, M. Kawasaki, Y. Tokura, K. Terakura, The anomalous Hall effect and magnetic monopoles in momentum space. *Science* **302**, 92–95 (2003).

46. D. Telesca, Y. Nie, J. I. Budnick, B. O. Wells, B. Sinkovic, Surface valence states and stoichiometry of non-superconducting and superconducting FeTe films. *Surf. Sci.* **606**, 1056–1061 (2012).

47. U. R. Singh, Preparation of magnetic tips for spin-polarized scanning tunneling microscopy on Fe$_{1+y}$Te. *Phys. Rev. B* **91** (2015).

48. J. Moon, Q. Zou, H. Zhang, O. M. J. van 't Erve, N. G. Combs, L. Li, C. H. Li, Magnetic field-induced spin nematic phase up to room temperature in epitaxial antiferromagnetic FeTe thin films grown by molecular beam epitaxy. *ACS Nano* **17**, 16886–16894 (2023).

49. D. Telesca, Y. Nie, J. I. Budnick, B. O. Wells, B. Sinkovic, Surface valence states and stoichiometry of non-superconducting and superconducting FeTe films. *Surf. Sci.* **606**, 1056–1061 (2012).

50. D. Telesca, Y. Nie, J. I. Budnick, B. O. Wells, B. Sinkovic, Impact of valence states on the superconductivity of iron telluride and iron selenide films with incorporated oxygen. *Phys. Rev. B* **85**, 214517 (2012).

51. S. Okamoto, A. Moreo, N. Nagaosa, S. S. P. Parkin, Berry curvature and intrinsic anomalous Hall effect in an antiferromagnet FeTe under magnetic field. private communication.

**Figures and Tables**

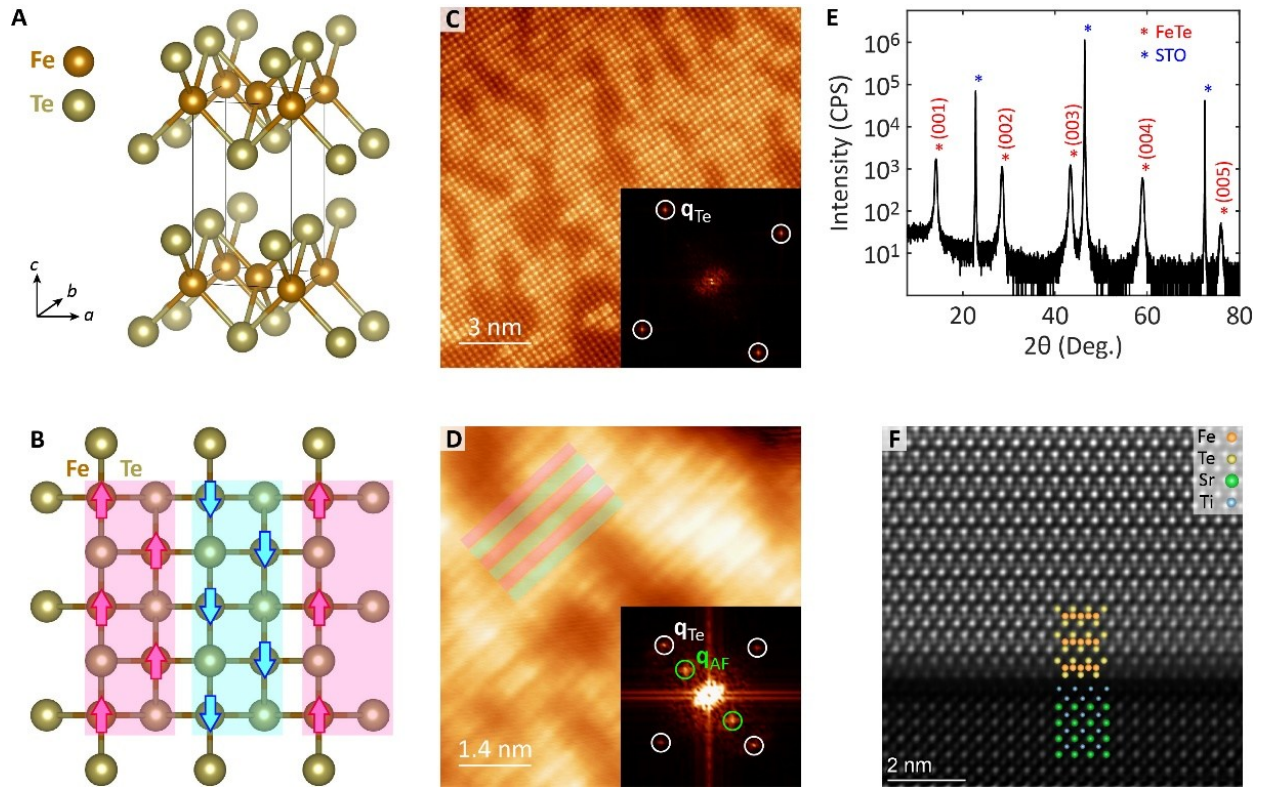

**Fig. 1. Structural characterizations of the FeTe/STO film.** (**A**) Schematic lattice structure of FeTe. (**B**) Schematic diagram of the bicollinear antiferromagnetic order in the ground state on the *a-b* plane of FeTe. The Fe spins are marked by red and blue arrows. (**C**) Atomically resolved STM topography of a FeTe film ($V$ = -100 mV, $I$ = 100 pA). Inset: FFT patterns of the topographic image. (**D**) STM topography acquired with a spin-polarized tip ($V$ = -100 mV, $I$ = 200 pA). Both the stripes of 2*a* spacing and the peaks $\mathbf{q}_{AF}$ = 1/2 $\mathbf{q}_{Te}$ in the FFT patterns (inset) indicate the bicollinear AF order in **b**. The blue and red stripes indicate the configuration of bicollinear AF order. (**E**) XRD patterns of the FeTe/STO film with evident (*00n*) brag peaks for both of FeTe and STO. (**F**) Cross sectional HAADF-STEM image near the interface between FeTe and STO, viewed along [010] direction. To clarify the interface structure, the color-coded atomic models are partially superimposed on the STEM data.

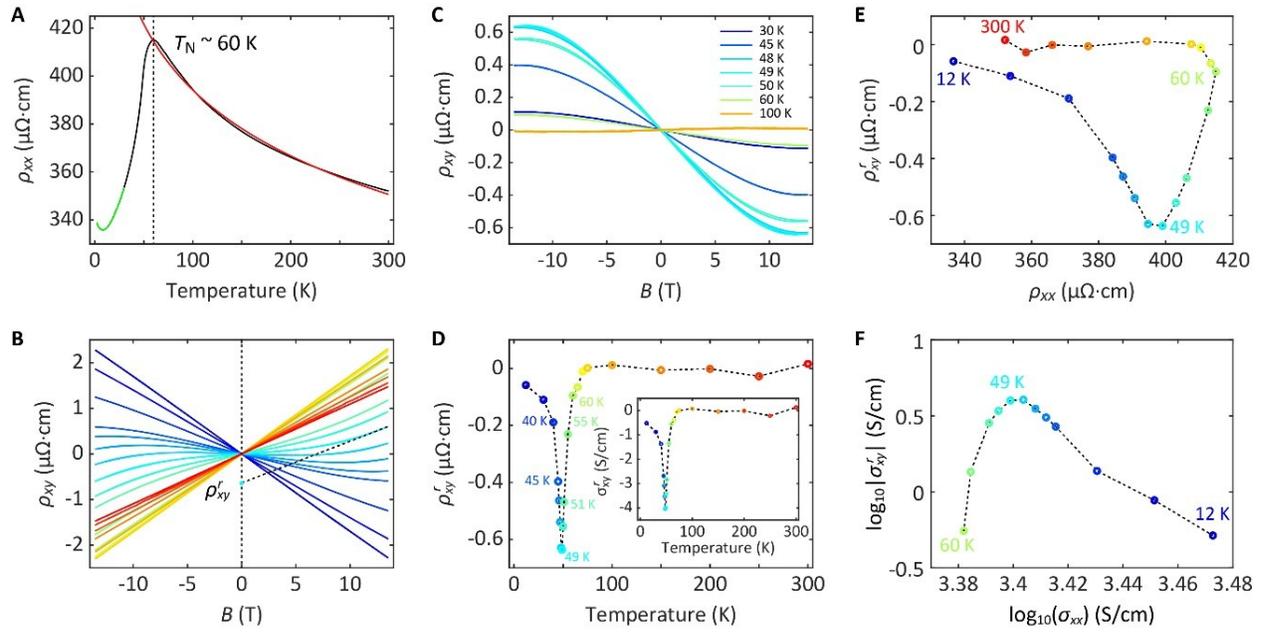

**Fig. 2. Transport properties and anomalous Hall effect in FeTe/STO.** (**A**) Longitudinal resistivity ($\rho_{xx}$) of the FeTe/STO film. The Néel temperature of the antiferromagnetic transition is identified as 60 K (vertical dashed line). The green curve represents a fit to the $\rho_{xx}(T)$ data for $T < 30$ K using the equation $\rho_{xx}(T) = a_1+b_1T^2+c_1\ln(T)+d_1T^5$, while the red curve shows a fit for $T > 65$ K with $\rho_{xx}(T) = a_2+b_2\ln(T)$. (**B**) Transverse resistivity ($\rho_{xy}$) as a function of magnetic field ($B$), measured at temperatures ranging from 12 K to 300 K. The temperatures are coded with colors, as indicated in (**D**). The slanted dashed line demonstrates a linear fit of $\rho_{xy}$-$B$ curve at high fields ($B > 12.5$ T), with the $y$-intercept defining the residual Hall resistivity ($\rho^r_{xy}$). (**C**) Field dependence of the residual transverse resistivity $\rho_{xy}$ after linear background subtraction. (**D**) Temperature dependence of $\rho^r_{xy}$ in FeTe/STO. Each data point is obtained from the linear fit of the corresponding color-coded $\rho_{xy}$-$B$ curve in (**B**). (**E**) Scaling relation between $\rho^r_{xy}$ and $\rho_{xx}$. (**F**) Logarithm scaling law between the absolute value of residual Hall conductivity $|\sigma^r_{xy}|$ and $\sigma_{xx}$.

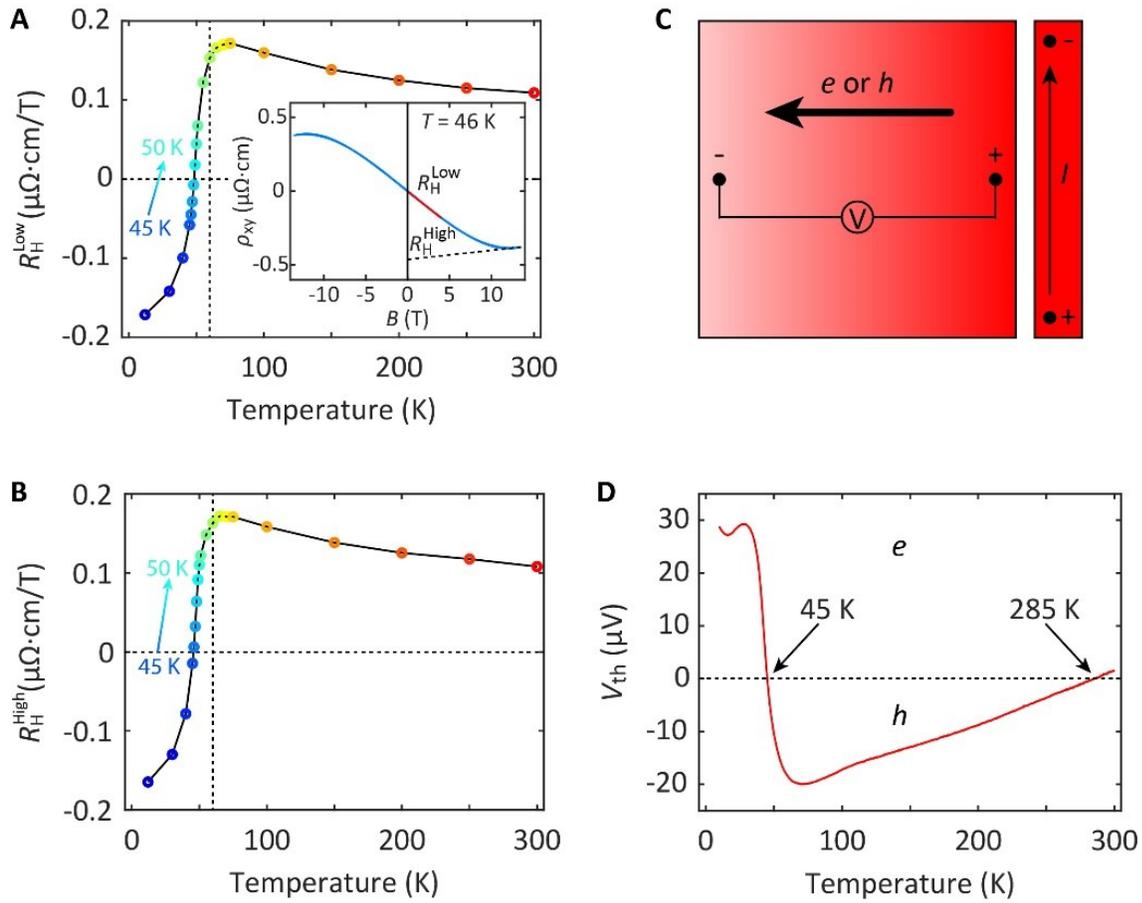

**Fig. 3. Thermal voltage and Hall coefficients of FeTe/STO.** Temperature dependence of the low-field Hall coefficient ($R_H^{Low}$) (**A**) and the high-field Hall coefficient ($R_H^{High}$) (**B**) extracted from the corresponding fitting ranges. Inset: Method for determining the Hall coefficients from the $\rho_{xy}$-$B$ curve. $R_H^{Low}$ is extracted through linear fitting in the low-field range ($B < 4$ T), while $R_H^{High}$ is obtained from linear fitting in the high-field range (B > 12.5 T). (**C**), Schematic diagram of the thermoelectric transport measurement setup. The FeTe film is etched to create an isolate current heating section (right) and a carrier drift region (left). (**D**), Thermal voltage $V_{th}$ of FeTe as a function of temperature. Positive and negative voltage values indicate electron and hole-dominated transport, respectively, with carrier-type transitions at approximately 45 K and 285 K.

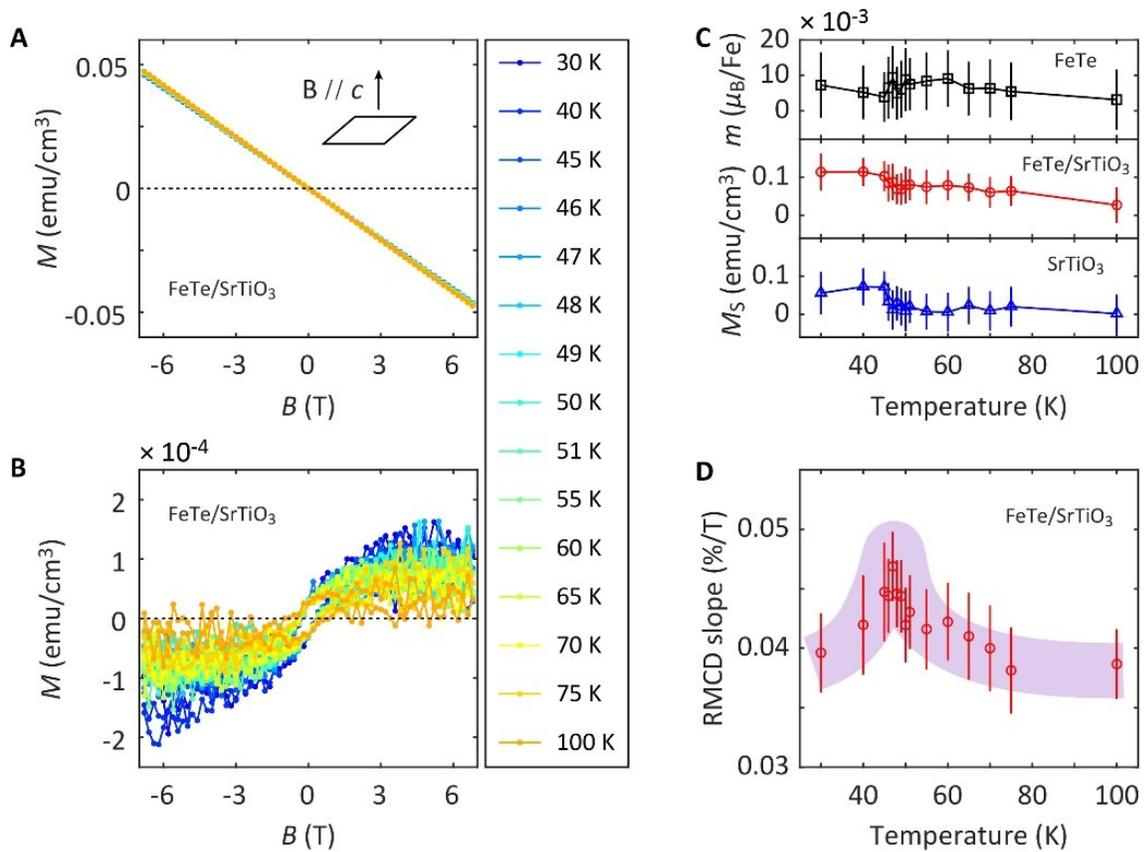

**Fig. 4. Magnetic moment of FeTe/STO.** (**A**) Field-dependent magnetization ($M$) measured by a magnetometer with an out-of-plane magnetic field ($B \parallel c$) at various temperatures. (**B**) Magnetization ($M$) vs $B$ with a linear background subtracted. The temperature-color correspondence in the legend is shared by (**A** and **B**). (**C**) Saturation magnetization ($M_S$) for $B \parallel c$ as a function of temperature for the STO substrate (bottom panel), FeTe/STO (middle panel) and the difference between them (top panel), extracted from the intercepts of linear fits to the high-field range of the $M(B)$ curves. (**D**) Temperature-dependent change in RMCD per unit field for FeTe/STO by linear fitting. The purple shadow is the guide for eyes.

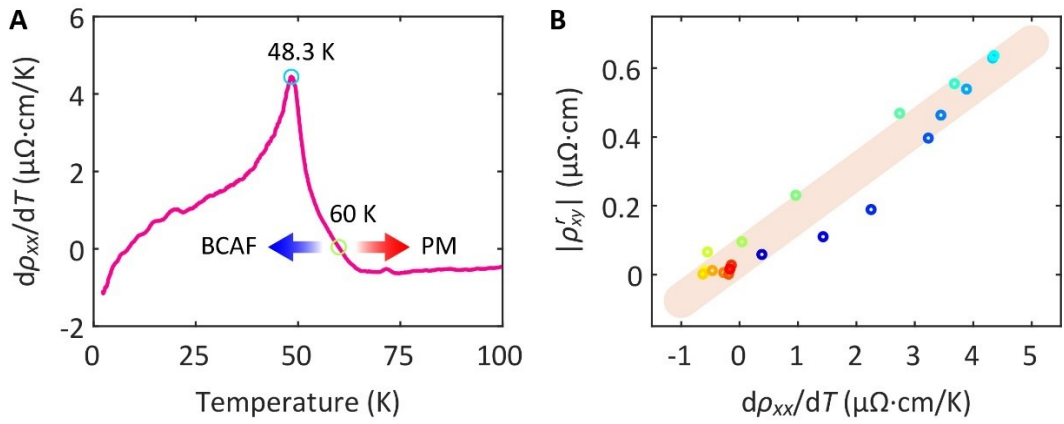

**Fig. 5. Correlation between residual Hall resistivity and longitudinal resistivity. (A)** Temperature dependence of the first derivative of the longitudinal resistivity, $d\rho_{xx}/dT$. The maximum at 48.3 K coincides with the temperature at which the residual Hall $\rho_{xy}^r$ reaches its peak. **(B)** $|\rho_{xy}^r|$ plotted as a function of $d\rho_{xx}/dT$ at different temperatures. The data show a linear trend where the shadow represents a linear fit.